\documentclass{PoS}
\usepackage{amsmath}
\usepackage{amsfonts}
\usepackage{amssymb}
\usepackage{graphicx}
\usepackage{caption}
\usepackage{subcaption}

\title{2p-2h excitations in neutrino scattering: 
angular distribution and frozen approximation}

\ShortTitle{2p-2h excitations in neutrino scattering}

\author{\speaker{I. Ruiz Simo}
\thanks{Special thanks to L. Alvarez-Ruso for invitation},$^a$
C. Albertus,$^a$ J.E. Amaro,$^a$
M.B. Barbaro,$^b$ J.A. Caballero,$^c$
and T.W. Donnelly$^d$\\
\llap{$^a$}  Departamento de F\'isica At\'omica,
Molecular y Nuclear, and Instituto de F\'isica 
Te\'orica y Computacional Carlos I, Universidad de Granada,
Granada 18071, Spain\\
\llap{$^b$}  Dipartimento di Fisica, Universit\`{a} di Torino
and INFN, Sezione di Torino, Via Pietro Giuria 1, 10125 Torino, 
Italy\\
\llap{$^c$} Departamento de F\'isica At\'omica, Molecular y
Nuclear, Universidad de Sevilla, Apartado 1065, 41080 Sevilla,
Spain\\
\llap{$^d$} Center for Theoretical Physics, Laboratory 
for Nuclear Science and Department of Physics, Massachusetts 
Institute of Technology, Cambridge, Massachusetts 02139, USA\\
E-mail: \email{ruizsig@ugr.es}}

\abstract{We study the phase-space 
dependence of 2p-2h excitations in neutrino
scattering using the relativistic Fermi gas model \cite{Rui14}.  We
follow a similar approach to Refs. \cite{Pace03,Ama10}, but focusing
in the phase-space properties, comparing with the non-relativistic
model of \cite{Van81}. A careful mathematical analysis of the angular
distribution function for the outgoing nucleons is performed.  Our
goals are to optimize the CPU time of the 7D integral to compute the
hadron tensor in neutrino scattering, and to conciliate the
different relativistic and non relativistic models by describing
general properties independently of the two-body current.  For some
emission angles the angular distribution becomes infinite in the Lab
system, and we derive a method to integrate analytically around the
divergence. Our results show that the frozen approximation, obtained by
neglecting the momenta of the two initial nucleons inside the integral
of the hadron tensor, reproduces fairly the exact response
functions for constant current matrix elements.}

\FullConference{16th International Workshop on Neutrino Factories and Future Neutrino Beam Facilities\\
		 25 -30 August, 2014\\
		 University of Glasgow, United Kingdom}

\begin{document}

\section{Introduction}
The analysis of neutrino oscillation experiments
requires having under control all the nuclear
effects, which are inherent to any $\nu$-nucleus
scattering event. These effects play a major role
in modifying the free $\nu$-nucleon cross section,
and a good understanding of them is clearly mandatory
in order to reduce the systematic uncertainty of the models
employed in the experimental analysis.\\
There is strong theoretical evidence \cite{Martini:2009uj,
Nieves:2011pp,Amaro:2010sd} about a significant
contribution from multi-nucleon knockout to the inclusive
Charged Current (CC) cross section around and above the
quasielastic (QE) peak region.\\
There are, at least, three different microscopic models
\cite{Martini:2009uj,Nieves:2011pp,Amaro:2010sd,Amaro:2011aa}
which are based in the relativistic Fermi Gas. But they differ 
from each other in several assumptions and different nuclear
ingredients for the interaction. Therefore, it is really 
difficult to disentangle the origin of any discrepancy in
the final results.\\
Other models \cite{Lalakulich:2012ac,Sobczyk:2012ms}
are based on the available phase space just 
assuming a constant transition matrix element and fitting 
it to the experimental cross section.\\
Finally, our goal is to reduce the computational time
needed in this kind of calculations, or at least 
to establish what assumptions previously made by other
authors are really good enough in order to estimate 
accurately and fast the contribution of these
multi-nucleon processes to the inclusive channel.

\section{2p-2h phase space in the Relativistic Fermi
Gas model}
The hadron tensor for the 2p-2h channel in a fully
relativistic framework is given by:
 \begin{eqnarray}
  W^{\mu\nu}_{2p2h}&=&\frac{V}{(2\pi)^9}\int d^3p^\prime_1\,
  d^3p^\prime_2\, d^3h_1\, d^3h_2\, \frac{m^4_N}{E_1E_2E^\prime_1E^\prime_2}
  r^{\mu\nu}(\mathbf{p}^\prime_1,\mathbf{p}^\prime_2,\mathbf{h}_1,
  \mathbf{h}_2) \delta(E^\prime_1+E^\prime_2-E_1-E_2-\omega)
 \nonumber\\ 
 &&\Theta(p^\prime_1,p^\prime_2,h_1,h_2)
  \delta^3(\mathbf{p}^\prime_1+
  \mathbf{p}^\prime_2-\mathbf{h}_1-\mathbf{h}_2-\mathbf{q})
  \label{hadron_tensor}
 \end{eqnarray}
 where $m_N$ is the nucleon mass, $V$ is the volume of the system
 and we have defined the product of step functions,
  \begin{equation}
  \Theta(p^\prime_1,p^\prime_2,h_1,h_2)\equiv\theta(p^\prime_1-k_F)
  \theta(p^\prime_2-k_F)\theta(k_F-h_1)\theta(k_F-h_2)
 \end{equation}
 which encodes the nuclear model.\\
 Finally, the function 
 $r^{\mu\nu}(\mathbf{p}^\prime_1,\mathbf{p}^\prime_2,\mathbf{h}_1,
  \mathbf{h}_2)$ is the elementary ``hadron'' tensor for
  the transition of a nucleon pair with given initial $(\mathbf{h}_1,
  \mathbf{h}_2)$ and final $(\mathbf{p}^\prime_1,\mathbf{p}^\prime_2)$
  momenta, summed up over spin and isospin, given schematically in
  terms of the antisymmetrized two-body currents by:
  \begin{equation}
   r^{\mu\nu}(\mathbf{p}^\prime_1,\mathbf{p}^\prime_2,\mathbf{h}_1,
  \mathbf{h}_2)=\frac14\sum_{\sigma,\tau}
  j^\mu(1^\prime,2^\prime;1,2)^{*}_A j^\nu(1^\prime,2^\prime;1,2)_A
  \end{equation}
The above multidimensional integral (\ref{hadron_tensor}) 
can be done either numerically or its dimensions can be
further reduced under some approximations \cite{Van81,
Alberico:1983zg,Gil:1997bm}. We do not know exactly
the origin of the discrepancies between the available models.
These could be due to different two-body currents, 
to local Fermi Gas model rather than global 
one, or other effects. It is difficult 
to make any concluding assessment 
right now, but all those models should agree at the level
of phase space function $F(\omega,q)$, obtained assuming a
constant $r^{\mu\nu}$:
  \begin{eqnarray}
  F(\omega,q)\equiv\int d^3h_1\, d^3h_2\, d^3p^\prime_1\,
  \frac{m^4_N}{E_1E_2E^\prime_1E^\prime_2}
  \Theta(p^\prime_1,p^\prime_2,h_1,h_2)
  \delta(E^\prime_1+E^\prime_2-E_1-E_2-\omega)
 \end{eqnarray}
  where $r^{\mu\nu}=1$ and $\mathbf{p}^\prime_2=\mathbf{h}_1+
  \mathbf{h}_2+\mathbf{q}-\mathbf{p}^\prime_1$ integrating out the 
  delta function of momentum conservation.\\
  The remaining delta function enables analytical integration
  over the modulus of $\mathbf{p}^\prime_1$:
\begin{eqnarray}
 F(\omega,q)=2\pi \int d^3h_1\, d^3h_2\, d\theta^\prime_1\,
 \sin\theta^\prime_1 \frac{m^2_N}{E_1E_2}
 \sum_{\alpha=\pm}\frac{m^2_N p^{\prime2}_1}{
 |E^\prime_2 p^\prime_1-E^\prime_1\mathbf{p}^\prime_2\cdot
 \hat{\mathbf{p}}^\prime_1|}\Theta(p^\prime_1,p^\prime_2,h_1,h_2)\Bigg|_{
 p^\prime_1=p^{\prime(\alpha)}_1}
\end{eqnarray}
and the sum inside the integral sign runs over the two solutions 
$p^{\prime(\pm)}_1$ of the energy conservation delta function
(see appendix C on Ref. \cite{Rui14}).

\subsection{Frozen nucleon approximation}
The frozen nucleon approximation is just a particular
case of the mean-value theorem in several variables
\begin{eqnarray}
\int_{a}^{b} f(x)dx&=&f(c)(b-a) \quad \text{with} \quad 
c\in\left[a,b\right]\\
\int_{\mathcal{V}}f(\mathbf{r})d^n\mathbf{r}&=&f(\mathbf{c})
\int_{\mathcal{V}}d^n\mathbf{r}=f(\mathbf{c})
\mathcal{V}\quad \text{with} \quad 
\mathbf{c}\in\mathcal{V}\label{theorem_n_dimensions}
\end{eqnarray}
In our case we are going to skip the two 3D integrations
over the holes momenta $(\mathbf{h}_1,\mathbf{h}_2)$ by
fixing them to some, in principle unknown, values inside
the Fermi sphere, while keeping the integration over the
emission angle and constraining it by energy conservation.
Of course, the obvious particularization of the mean-value
theorem (\ref{theorem_n_dimensions}) is:
\begin{eqnarray}
 F(\omega,q)&=&\int d^3h_1 d^3h_2 d^3p^\prime_1\; f(\mathbf{h}_1,
 \mathbf{h}_2,\mathbf{p}^\prime_1)=
 \left(\frac43\pi k^3_F
 \right)^2 \int d^3p^\prime_1\; f(\langle\mathbf{h}_1\rangle,
 \langle\mathbf{h}_2\rangle,\mathbf{p}^\prime_1)
\end{eqnarray}
where $(\langle\mathbf{h}_1\rangle,
 \langle\mathbf{h}_2\rangle)$ are the 
 two unknown hole momenta inside
 the Fermi sphere.\\
 Up to now the whole discussion is exact. What makes
 the difference between an excellent approximation
 to the true result or a poor one is just
 the selection of the ``average''
 hole momenta $(\langle\mathbf{h}_1\rangle,
 \langle\mathbf{h}_2\rangle)$. Our choice to call it
 \emph{frozen nucleon approximation} will be
 $(\vec{0},\vec{0})$. There are mainly two arguments
 which favor this choice:
 \begin{itemize}
  \item For high $q>>k_F>h_i$, one can assume both initial
  nucleons at rest.
  \item If one does not assume the above statement, one has to
  determine the average angles for the two holes. And
  this cannot be done without computing the full 7D integral
  (see section \ref{sec:other_configurations}).
 \end{itemize}
 Now, we can define the \emph{frozen approximation}
 phase-space function $\overline{F}(\omega,q)$:
 \begin{equation}
  \overline{F}(\omega,q)=\left(\frac43\pi k^3_F\right)^2
  \int d^3p^\prime_1\,\delta(E^\prime_1+E^\prime_2
  -\omega-2m_N)\,\Theta(p^\prime_1,p^\prime_2,0,0)
  \frac{m^2_N}{E^\prime_1 E^\prime_2}\label{Fbar}
 \end{equation}
 where now $\mathbf{p}^\prime_2=\mathbf{q}-\mathbf{p}^\prime_1$.\\
 The above phase-space function (\ref{Fbar}) allows us to
 define the angular distribution in the \emph{frozen
 approximation}:
 \begin{eqnarray}
  \overline{F}(\omega,q)&=&\left(\frac43\pi k^3_F\right)^2
  \,2\pi \int^{\pi}_{0} d\theta^\prime_1\,\Phi(\theta^\prime_1)\\
\Phi(\theta^\prime_1)&=&\sin\theta^\prime_1\int dp^\prime_1\,
  p^{\prime\,2}_1\,\delta(E^\prime_1+E^\prime_2
  -\omega-2m_N)\,\Theta(p^\prime_1,p^\prime_2,0,0)
  \frac{m^2_N}{E^\prime_1 E^\prime_2}
 \end{eqnarray}
 \vspace{-2cm}
 \begin{figure}[h]
\centering
\begin{subfigure}{.5\textwidth}
  \hspace{-2.5cm}
  \includegraphics[height=16cm,width=12cm]{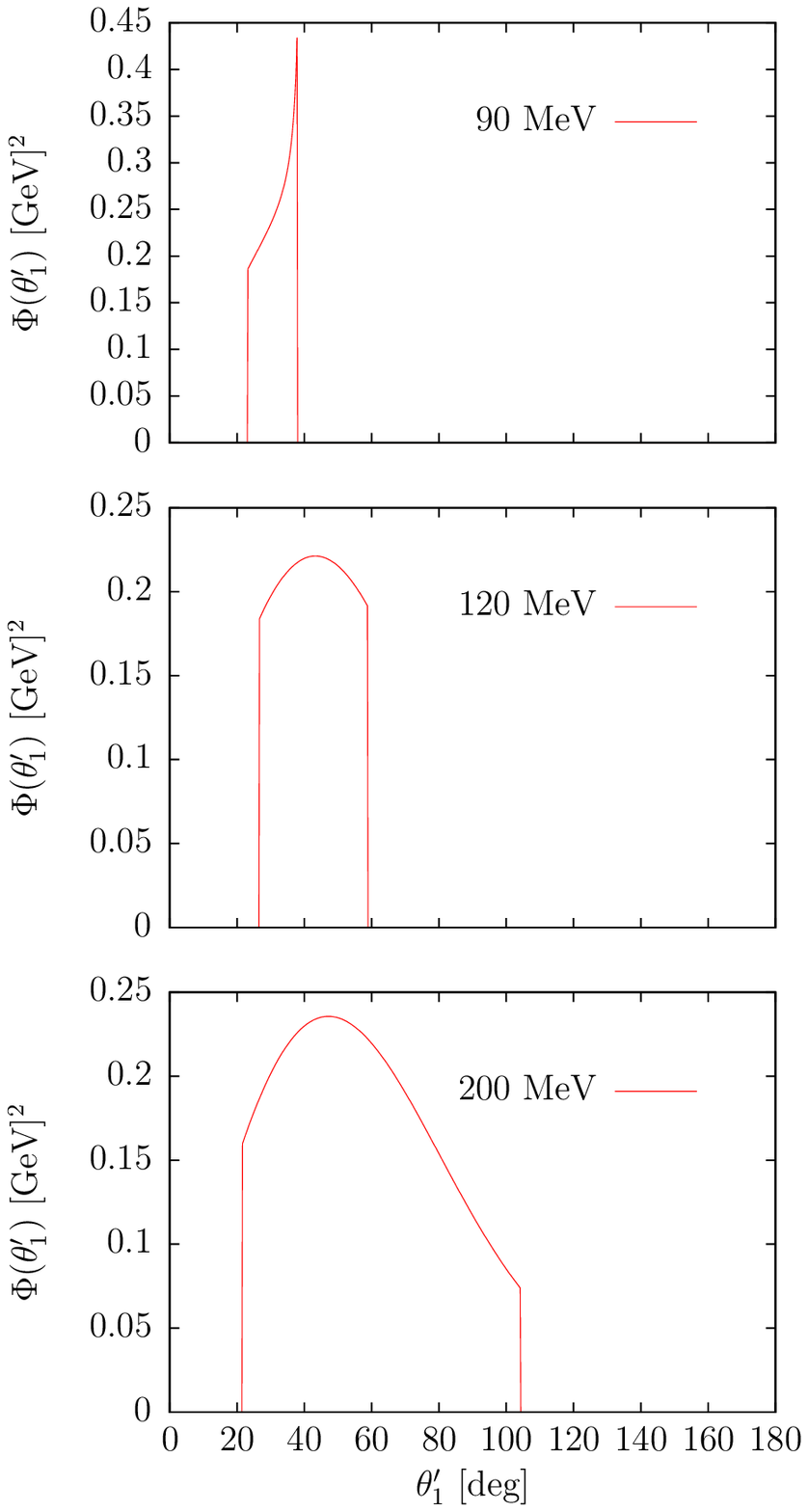}
 \vspace{-6cm}
  \caption{}\label{fig:angular_distribution_500mev}
 \end{subfigure}%
\begin{subfigure}{.5\textwidth}
  \hspace{-2.25cm}
  \includegraphics[height=16cm,width=12cm]{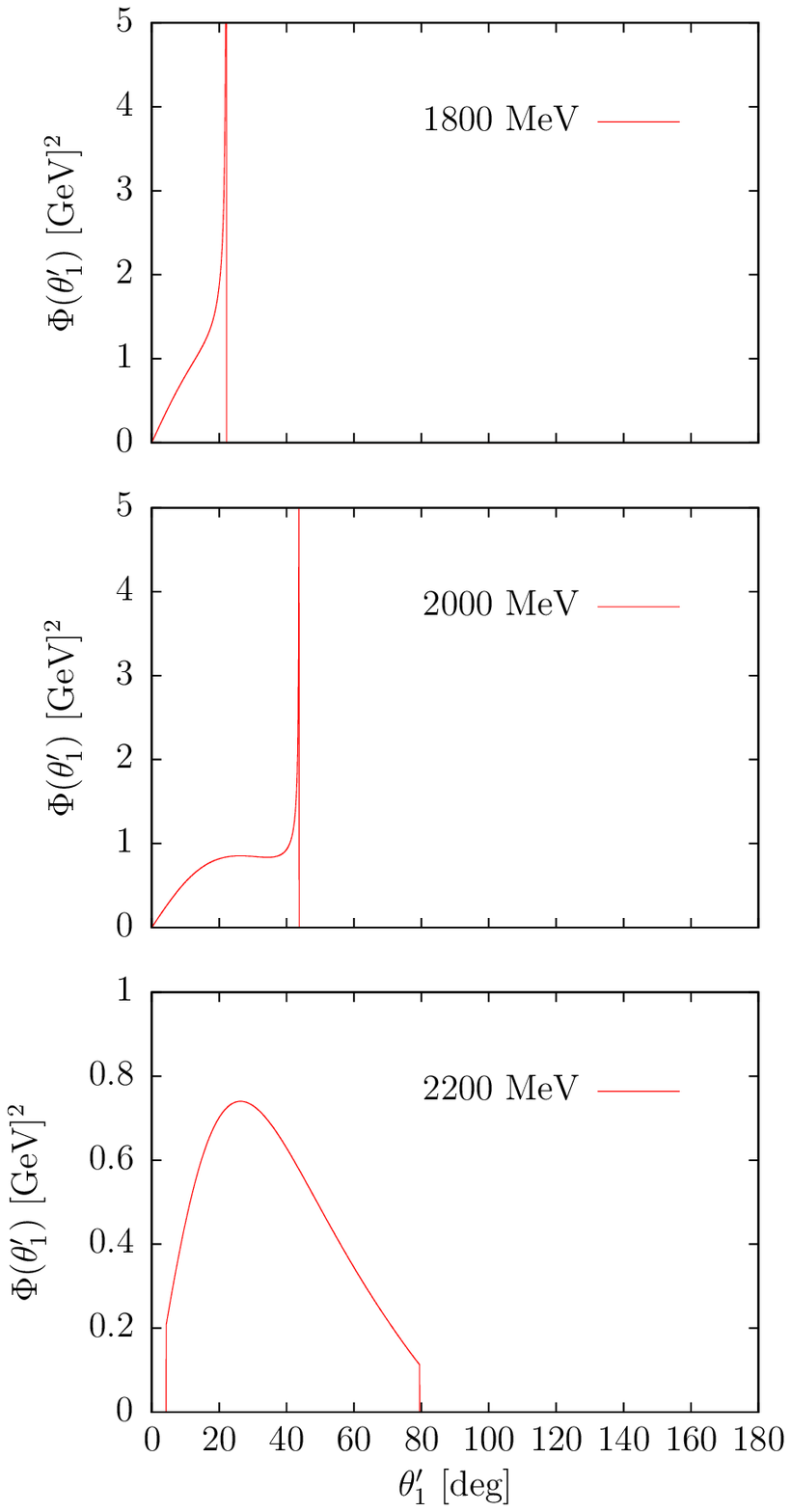}
  \vspace{-6cm}
  \caption{}\label{fig:angular_distribution_3gev}
  \end{subfigure}
\caption{Angular distributions $\Phi(\theta^\prime_1)$
  for $q=500$ MeV/c (left panel) and for $q=3$ GeV/c
  (right panel) in the frozen nucleon approximation, for
  three different values of $\omega$ as a function of the 
  emission angle $\theta^\prime_1$.}
\label{fig:angular_distribution}
\end{figure}
\\
The problem stands on the divergence of the angular
distribution for some kinematics. This can be seen in 
some of the panels of figure \ref{fig:angular_distribution}
for the frozen nucleon approximation. For other values of 
the holes momenta, the position of the pole is 
different, but it is still present. This makes 
crucial to determine analytically the angular
interval where the integration has to be performed.
This was done explicitly on section VI of Ref.
\cite{Rui14}.\\

 \begin{figure}
\centering
\begin{subfigure}{.5\textwidth}
  \hspace{-2.5cm}
  \includegraphics[height=16cm,width=12cm]{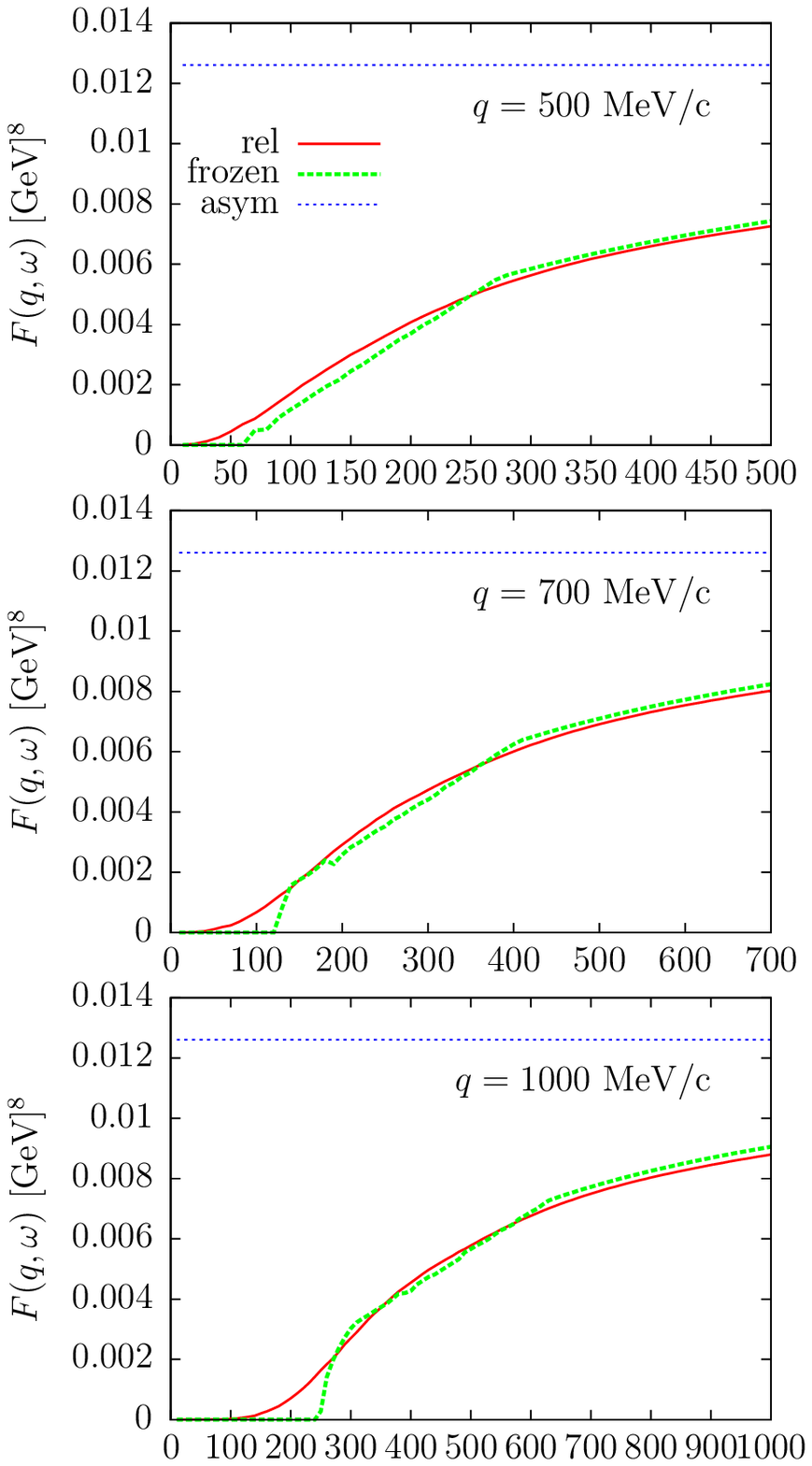}
    \vspace{-6cm}
    \caption{}\label{fig:comparison_lowq}
   \end{subfigure}%
\begin{subfigure}{.5\textwidth}
  \hspace{-2.25cm}
  \includegraphics[height=16cm,width=12cm]{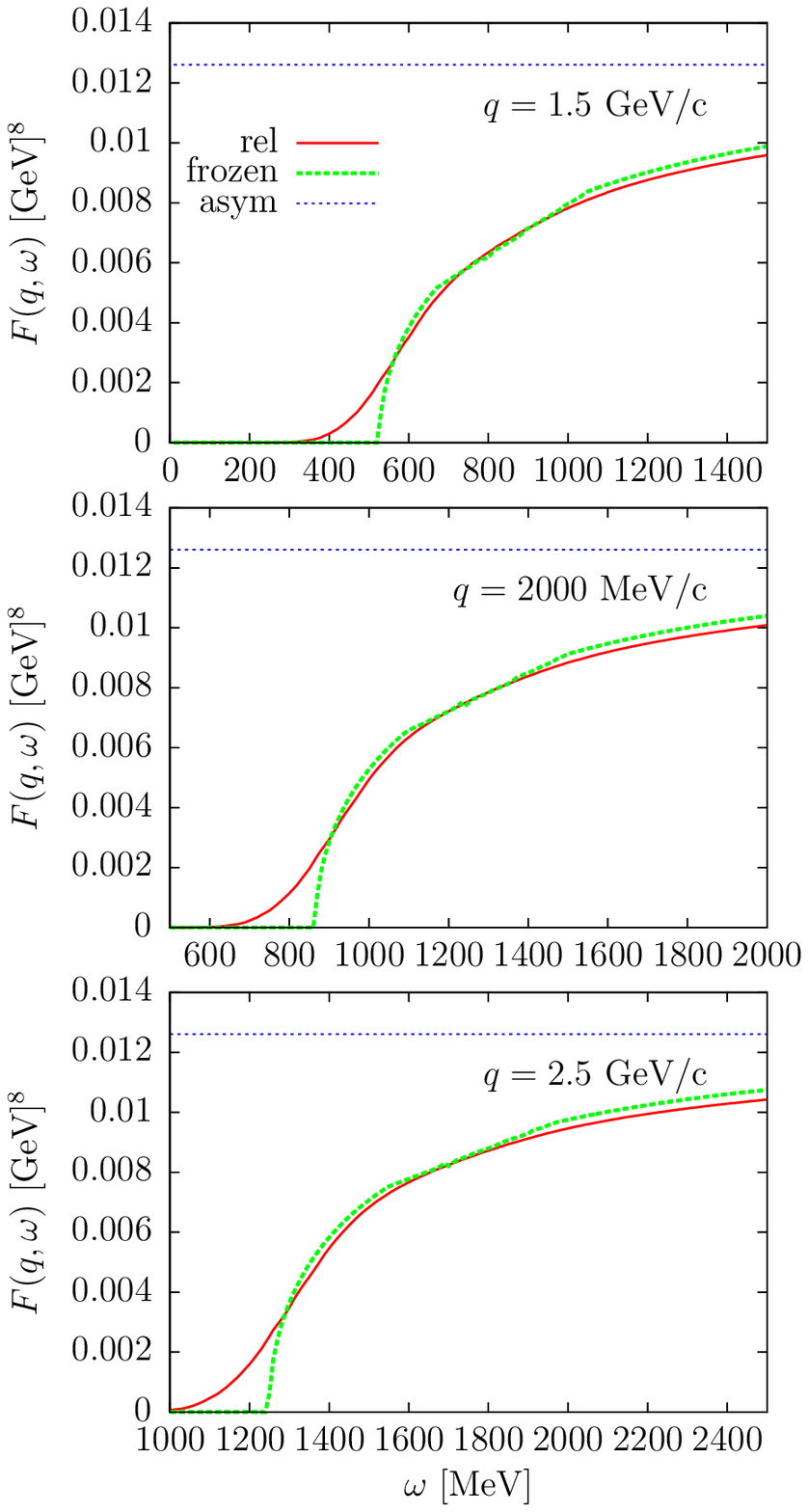}
  \vspace{-6cm}
  \caption{}\label{fig:comparison_highq}
  \end{subfigure}
  \caption{Comparison between frozen nucleon approximation 
  and full integral for low
    and intermediate momentum transfers (left panel) and 
    for high momentum transfers (right panel).}
    \label{fig:comparison_phase_space}
  \end{figure}
Once the problem of the divergence has been correctly
addressed, the results for the phase space function in
the frozen approximation can be compared with the full
integral in figure \ref{fig:comparison_phase_space}.
Here we can appreciate the quality of the approximated
result for a wide range of momentum transfers. The main 
discrepancies arise on the low energy-transfer region
in each plot. But the phase space function is 
well-reproduced in the rest of the interval and, furthermore,
we have skipped 6 additional integrals (over holes momenta), 
thus reducing significantly the computation time.
These results indicate that this approximation is
especially well-suited for Montecarlo
event generators, particularly if the goal is 
to estimate quite accurately the cross section for 
multi-nucleon knockout in the shortest time as possible.
\subsection{Other initial configurations}\label{sec:other_configurations}
\begin{figure}[h]
\centering
\begin{subfigure}{.5\textwidth}
  \hspace{-1.5cm}
  \vspace{5cm}
  \includegraphics[height=15cm,width=10.5cm]{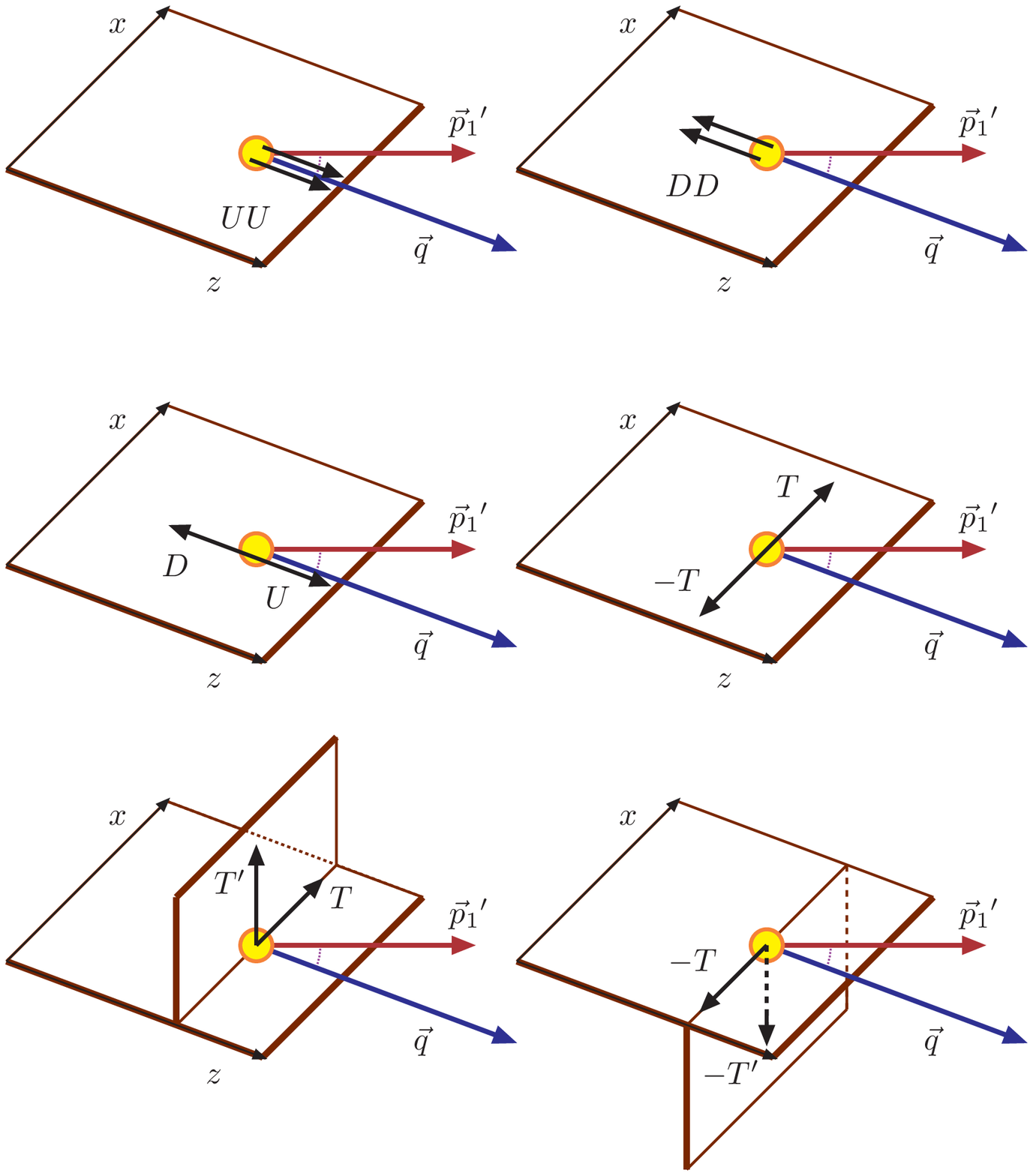}
    \vspace{-10cm}
    \caption{}\label{fig:configurations_a}
   \end{subfigure}%
\begin{subfigure}{.5\textwidth}
  \hspace{-1.25cm}
  \vspace{3cm}
  \includegraphics[height=16cm,width=12cm]{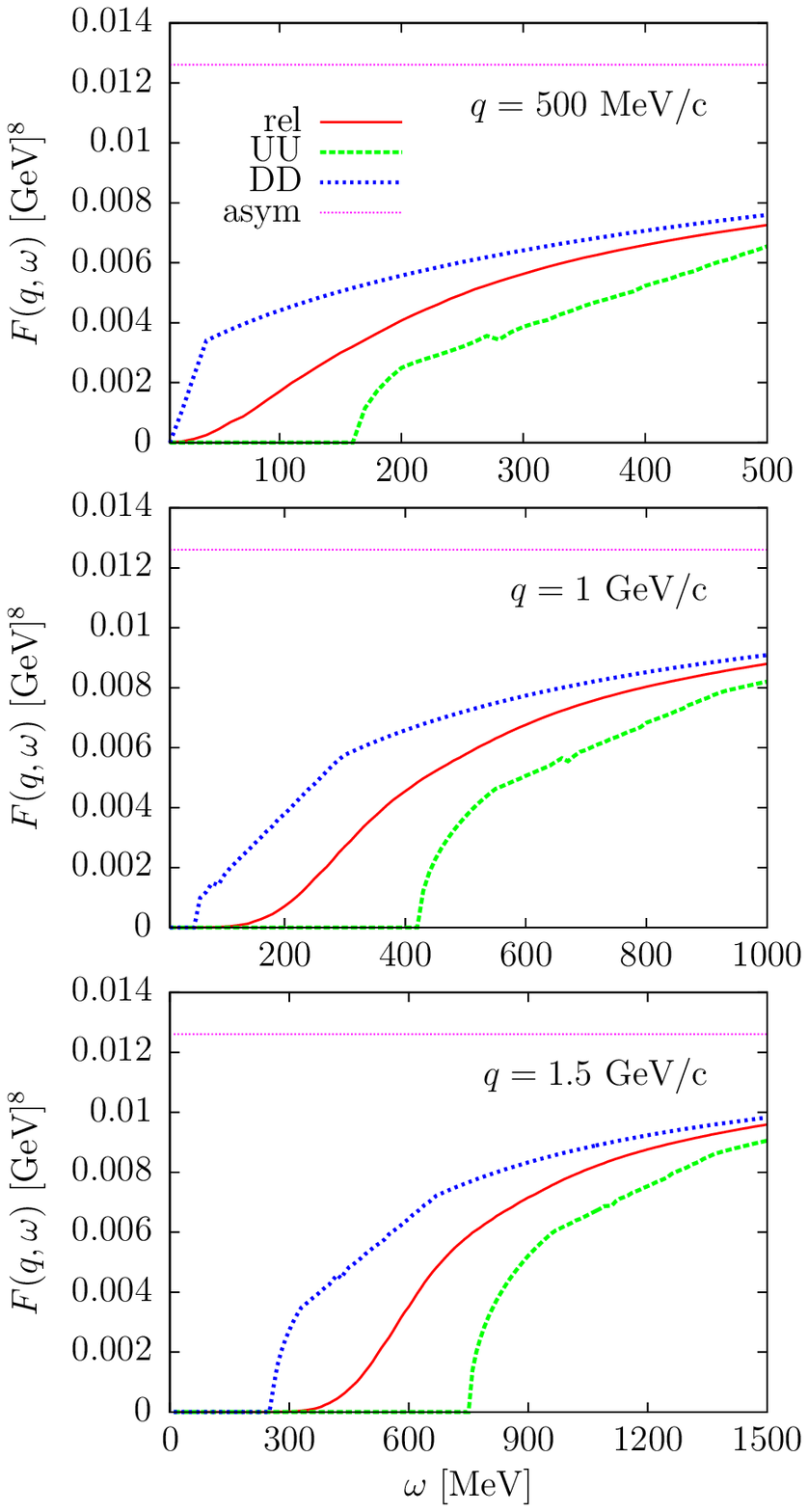}
  \vspace{-8.5cm}
  \caption{}\label{fig:configurations_b}
  \end{subfigure}
  \caption{Geometry employed for emission of a 
  pair of nucleons with initial parallel momenta (UU,DD),
    anti-parallel (UD,T-T) and perpendicular (TT',-T-T') on 
    the left panel. In the right panel a 
    comparison between 
    average momentum approximation and full
    integral is shown. Initial momenta are both 
    200 MeV/c.}\label{fig:configurations}
  \end{figure}

\begin{figure}
\centering
\begin{subfigure}{.5\textwidth}
  \hspace{-2.5cm}
  \includegraphics[height=16cm,width=12cm]{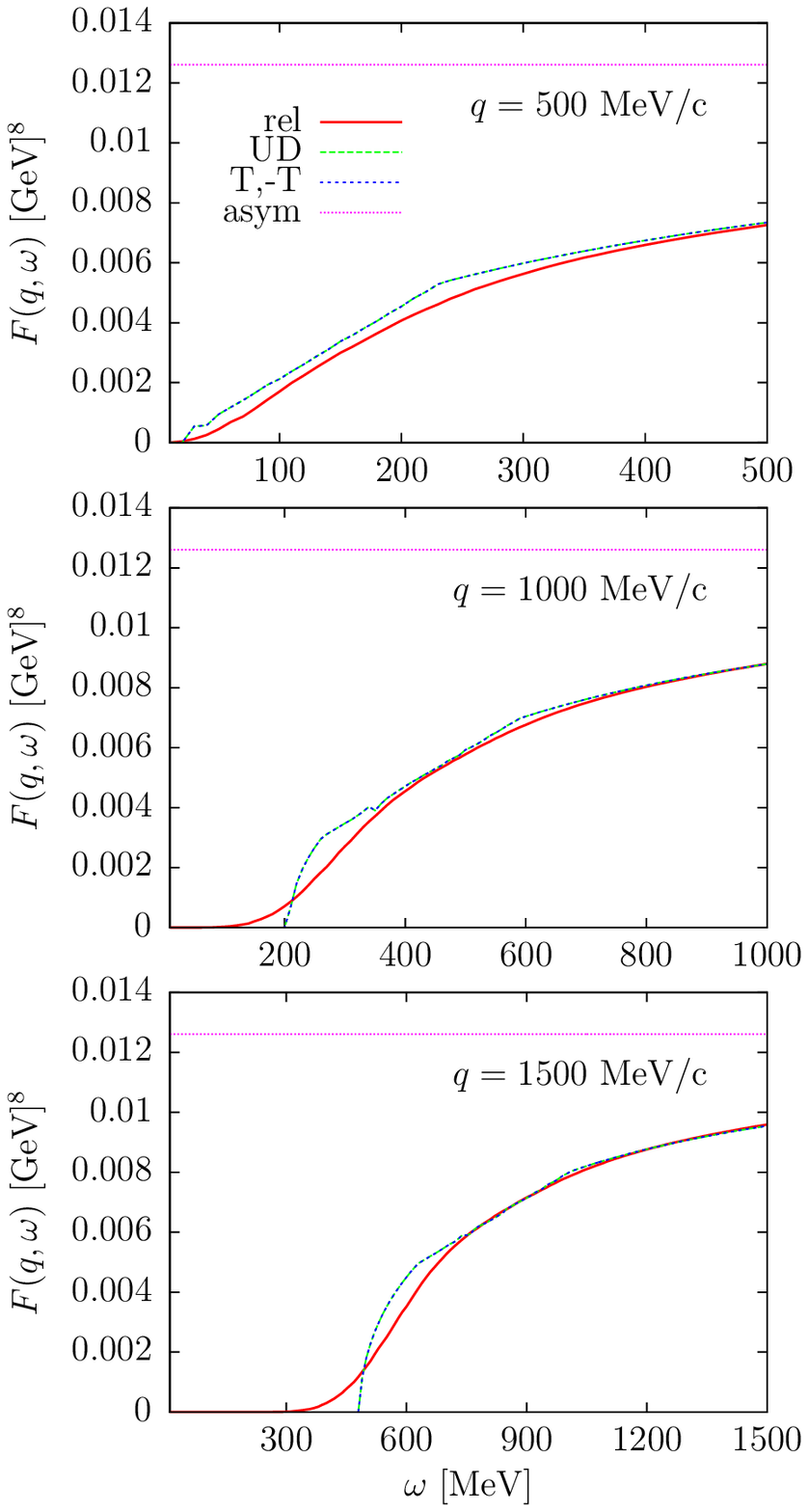}
    \vspace{-6cm}
    \caption{}\label{fig:comparison_configurations_zero}
   \end{subfigure}%
\begin{subfigure}{.5\textwidth}
  \hspace{-2.25cm}
  \includegraphics[height=16cm,width=12cm]{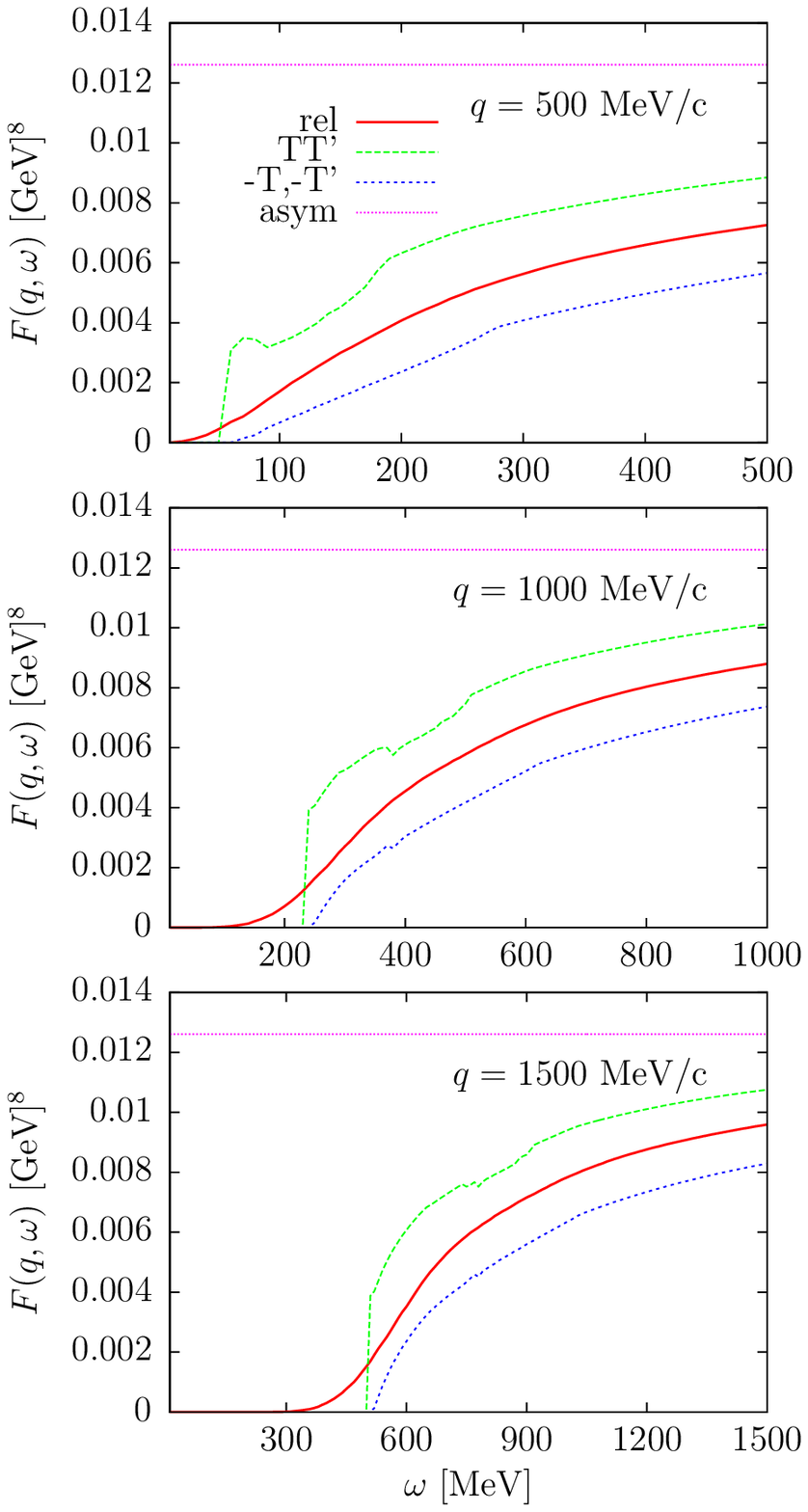}
  \vspace{-6cm}
  \caption{}\label{fig:comparison_configurations_orthogonal}
  \end{subfigure}
  \caption{Comparison between average momentum 
  approximation and full integral. In the left panel
  the initial momenta
  are both 200 MeV/c with total momentum of the pair 
  equal to zero. In the right one the initial momenta
  are both 200 MeV/c pointing in orthogonal directions.}
  \label{fig:comparison_configurations}
  \end{figure}

We have chosen up to now the 
\emph{frozen nucleon approximation}, but other initial
$(\langle\mathbf{h}_1\rangle,
 \langle\mathbf{h}_2\rangle)$ configurations could 
 have been selected as well. We have considered six
 configurations depicted in fig. \ref{fig:configurations_a}. 
 The configurations
 with total initial momentum of the pair equal to zero 
 (U,D or T,$-$T) would, in principle, yield a result very close to
 that of the \emph{frozen approximation}. This can be
 easily observed in figure 
 \ref{fig:comparison_configurations_zero}.
 
 In the first comparison, Fig. \ref{fig:configurations_b}, 
 we show the contribution
of two pairs of nucleons with the same momentum
$h_1=h_2=200$ MeV/c, and both parallel, pointing
upward (U) and downward (D) with respect to the z
axis, that is, the direction of $\mathbf{q}$. 
The contribution of the
UU configuration is smaller than average, while the DD
is larger. This is so because in the UU case the total
momentum $p^\prime$ in the final state is large. By momentum
conservation, the momenta $p^\prime_1$ and $p^\prime_2$
must also be large.
Therefore, these states need a large excitation energy, and
they start to contribute for high $\omega$ transfer. In the DD
configuration, the total momentum $p^\prime$ is smaller, so the final
momenta $p^\prime_1$ and $p^\prime_2$ can also be small, 
with small required excitation energy. Therefore, they start
to contribute at lower $\omega$.\\
In the example of Fig. \ref{fig:comparison_configurations_zero},
two anti-parallel configurations
are shown. In the UD case, one nucleon is moving upward
and the other downward the z axis with total momentum
zero of the pair. This situation is similar to that of a pair of
highly correlated nucleons with large relative momentum
\cite{Korover:2014dma}. Since the total momentum 
is zero, the final 2p-2h state
has total momentum q, exactly the same that it would have
in the \emph{frozen nucleon approximation}. Therefore, 
the contribution of this configuration is similar to the average.
The same conclusions can be drawn in the case of the
configuration $T$, $-T$, with one nucleon moving along the $x$
axis (transverse direction) and the other along $-x$ with
opposite momentum. The contribution of this pair is
exactly the same as that of the UD configuration in the
total phase-space function.\\
Finally, we show in Fig. 
\ref{fig:comparison_configurations_orthogonal} two 
intermediate cases that
are neither parallel nor anti-parallel configurations. They
consist of two pairs of transverse nucleons moving along
mutually perpendicular directions. In the first case, we
consider a $T$ nucleon and a second $T^\prime$ nucleon moving in
the y axis out of the scattering plane. The contribution of
the $TT^\prime$ pair is large, while the 
one of the opposite case, $-T$,
$-T^\prime$, is small. On the average, they are close to the total
result.

\section{Conclusions}
We have performed a detailed study of the two-particle-
two-hole phase-space function, which is proportional to the
nuclear two-particle emission response function for 
constant current matrix elements. Our final goal was
to obtain accurate enough results without calculating the 
7D integral. The frozen nucleon approximation (1D integral),
that is,
neglecting the momenta of the initial nucleons for high
momentum transfer, seems to be a quite promising approach
to reduce the computation time without missing significant
accuracy. The CPU time of the 7D
integral has been reduced significantly. 
We are presently working on an implementation of
the present method with a complete model of the MEC
operators.
\section{Acknowledgements}
This work was supported by DGI (Spain), Grants
No. FIS2011-24149 and No. FIS2011-28738-C02-01; by
the Junta de Andaluc\'ia (Grants No. FQM-225 and
No. FQM-160); by the Spanish Consolider-Ingenio 2010
programmed CPAN, in part (M. B. B.) by the INFN project
MANYBODY and in part (T. W. D.) by U.S. Department of
Energy under Cooperative Agreement No. DE-FC02-
94ER40818. C. A. is supported by a CPAN postdoctoral
contract.

\end{document}